\documentstyle[aps,prl,floats,epsfig]{revtex}

\begin{document}

\twocolumn
[\hsize\textwidth\columnwidth\hsize\csname
@twocolumnfalse\endcsname
\title{Perspectives on Finding the Neutrino Nature}
\author{M. Czakon$^a$, J. Gluza$^{a,b}$ and M. Zra{\l}ek$^a$}
\address{$^a$ Department of Field Theory and Particle Physics,
University of Silesia, Uniwersytecka 4, PL-40-007 Katowice, Poland \\
$^b$ Deutsches Elektronen-Synchrotron DESY, Zeuthen, Germany}
\date{\today}
\maketitle

\begin{abstract}
The possibility of determining the neutrino nature is considered in view of the most recent experimental observations.
The analysis combines schemes with three and four neutrinos.The data on oscillations is put together with that from the search
of neutrinoless double beta decay and results on tritium beta decay. All solar neutrino oscillation solutions are taken into
account. The sensitivity of the problem on future experimental bounds from GENIUS is studied. Dirac neutrinos are shown to
be unavoidable already at present in some schemes and the constraints will quickly become more stringent with future data.
The consequences of including bounds from Cosmology on the neutrino content of Hot Dark Matter are commented.
\end{abstract}

\vspace{0.5cm}]
\narrowtext


At present, particle physics seems to have a long awaited opportunity to discover
the path towards an extension of the Standard Model. This opportunity is given by
the recent confirmation of neutrino oscillations \cite{SK}. With the currently accumulated
experimental data, we have many hints on the size of neutrino masses and mixing angles. This
information imposes serious constraints on the structure of the model. Unfortunately there
remains a major unknown, which is the neutrino nature. As a neutral particle it can be
described either by means of the Dirac equation or by means of the Majorana equation \cite{zralek}.
This problem is of the utmost importance for model builders.

It has been proved many times, that with the current precision, the only experiment able
to show which of the two descriptions is valid, is the search for the neutrinoless double beta decay
$(\beta\beta)_{0\nu}$. Should this process be observed, we would be left with no other
choice but to acknowledge neutrinos to be Majorana particles. As this has not been the case,
we have at hand only a lower bound on the lifetime of the $(\beta\beta)_{0\nu}$ decaying nuclei.
This alone is not enough. However when supplied with the results of neutrino oscillation experiments
and the tritium beta decay mass bounds, interesting, but this time neutrino mass scheme
dependent conclusions can be drawn. The situation will 
undoubtedly improve, when the future GENIUS \cite{genius} experiment will furnish its first results.

In this paper, we combine the neutrino oscillation data, the tritium beta decay bounds and the
present and future GENIUS bounds on the $(\beta\beta)_{0\nu}$ lifetime (if their results are not
positive, since in the opposite case the situation is clear, as stated above) in order to derive
conditions which would define the neutrino as a Dirac particle. To this end we take into account
the possibility of having three or four light neutrinos. Therefore our conclusions will be valid
in either case, whether the LSND \cite{lsnd} results would eventually be ruled out or confirmed.

The novelty in respect to previous analyses \cite{our} is the consideration of the case of three and
four neutrinos together, supplemented with the most recent bounds on the mixing matrix elements and
the variation of the minimal neutrino mass $(m_\nu)_{min}$ in its full range. Furthermore
we do not assume that neutrinos are Majorana particles like in so many other works \cite{syf}.

\begin{figure}[t]
\epsfig{figure=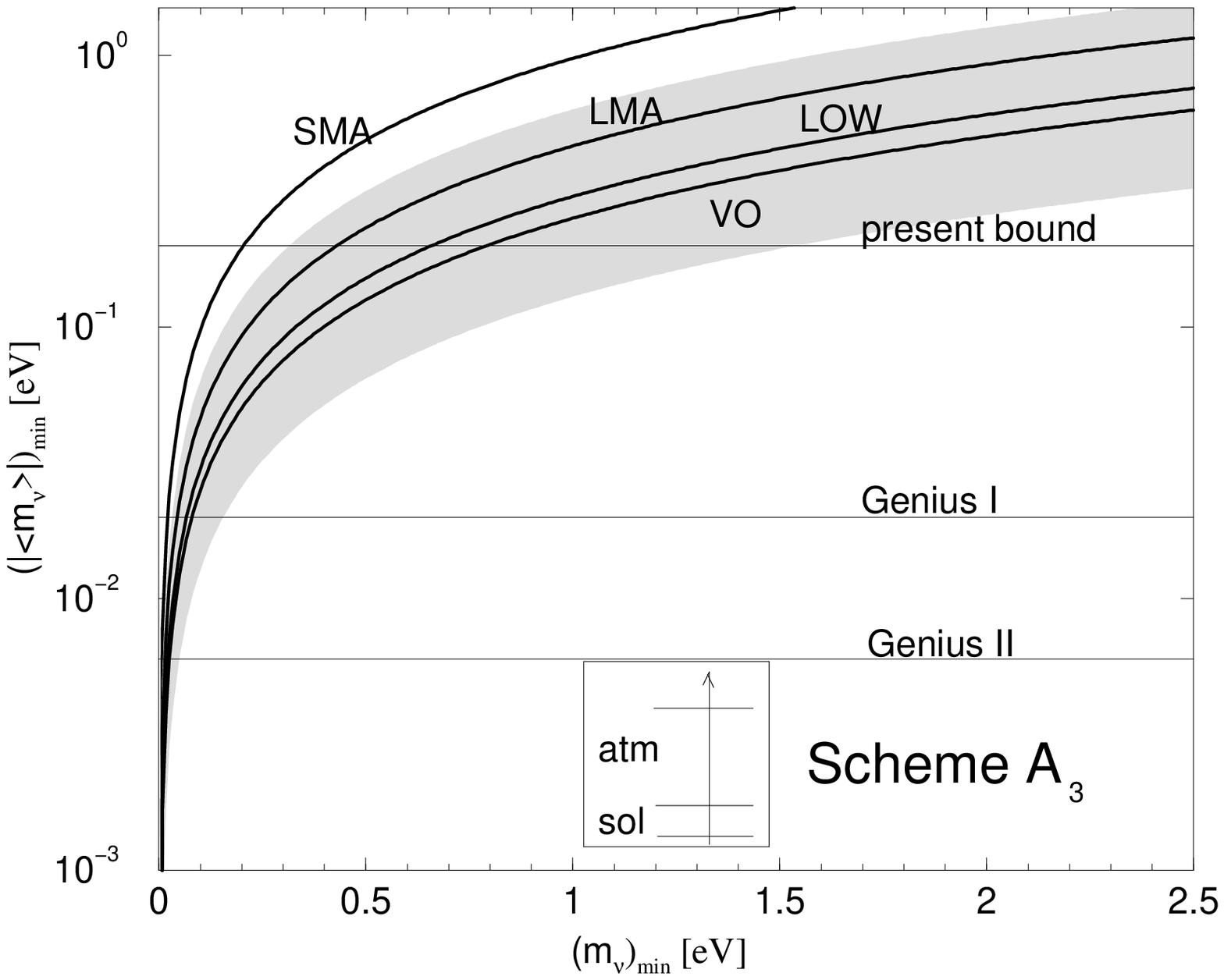, width=8.5cm} \\
\epsfig{figure=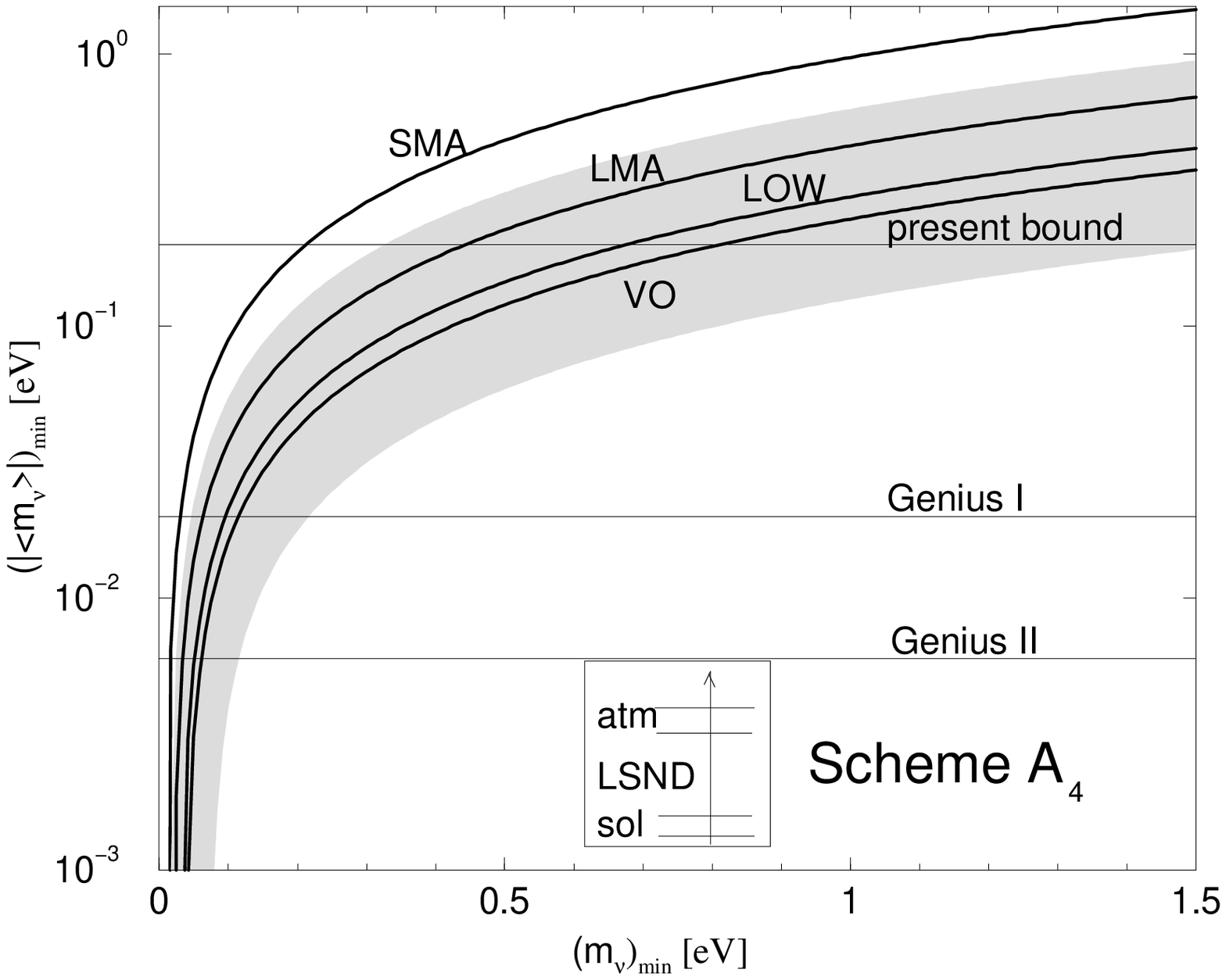, width=8.5cm}
\caption{Schemes $A_3$ and $A_4$. The curves for the different solar neutrino oscillation solutions were obtained with
the values of $\sin^2 2\theta_{sun}$ quoted in the text and $s^e_+=.01$ . The shaded region is given by the largest range in 
Table~\ref{zwei}, that is the range for the LMA solution. For the SMA solution the $95\%c.l.$ range (.001-.01) is described 
by one curve in the present scale.
\label{a3a4}}
\end{figure}

\begin{figure}[t]
\epsfig{figure=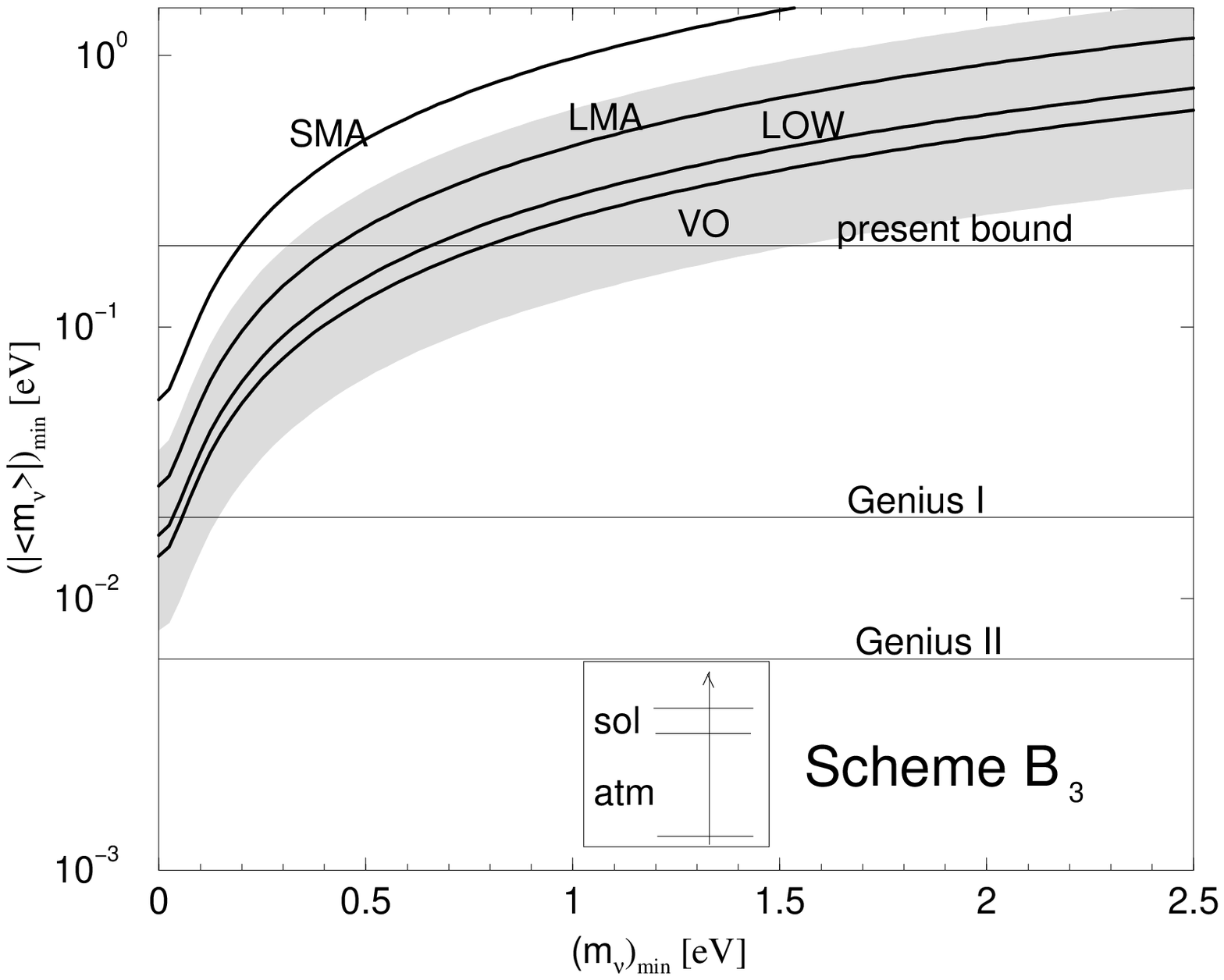, width=8.5cm} \\
\epsfig{figure=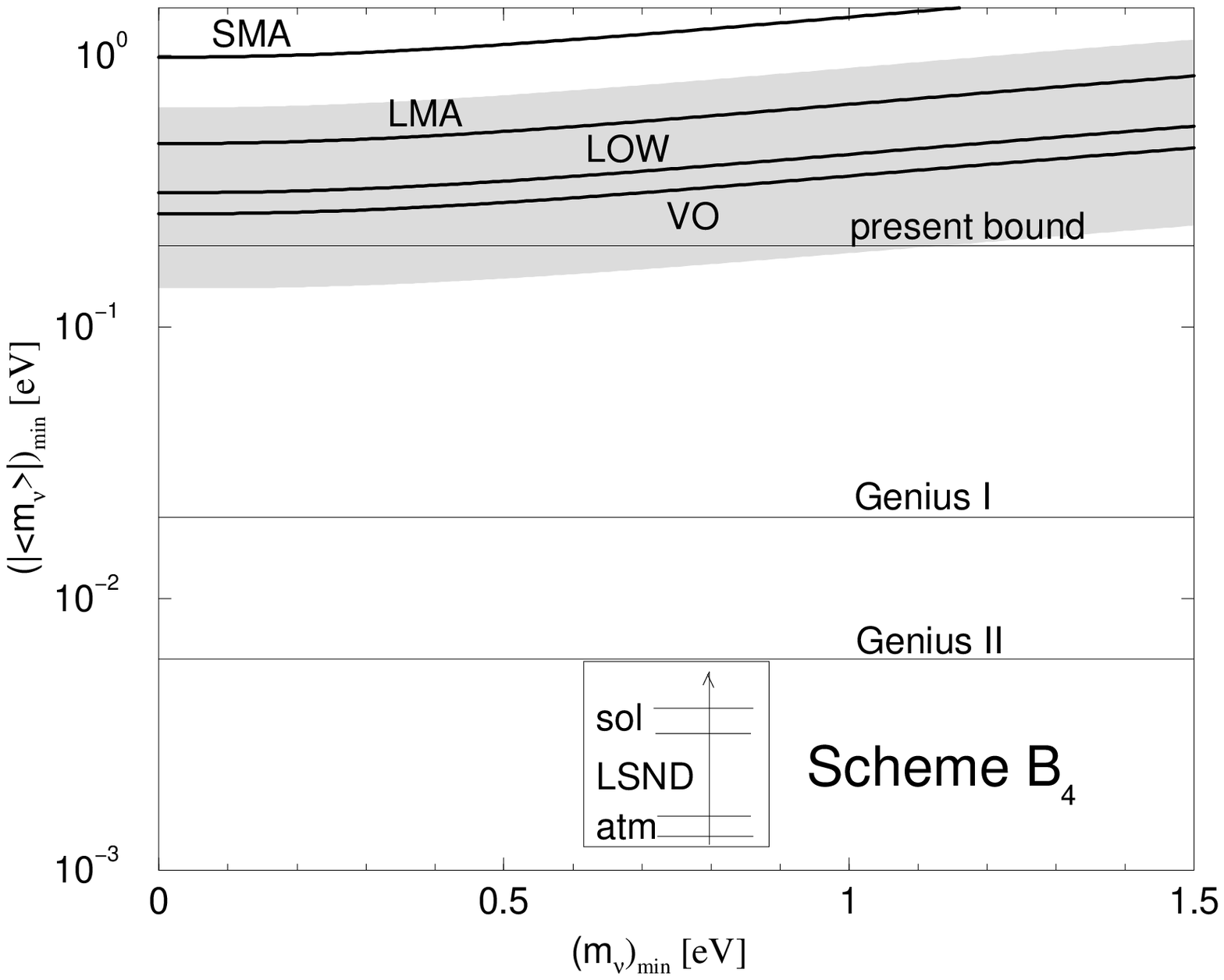, width=8.5cm}
\caption{Schemes $B_3$ and $B_4$. The curves were obtained with the same assumptions as in Fig.~\ref{a3a4}.
\label{b3b4}}
\end{figure}

The oscillation data is described in terms of differences of neutrino mass squares 
$\delta m^2_{ij} \equiv m^2_i-m^2_j$. Define a scheme of neutrino masses as an ordering
of these $\delta m^2$'s. For three neutrinos there are two such schemes, whereas for
four neutrinos there are six. In the latter case, four schemes can be rejected by means of
a combined analysis \cite{bilenky}. All four possibilities are shown in Fig.~\ref{a3a4} and Fig~\ref{b3b4}.
It should be clear that with the
strong ordering of $\delta m^2$'s ($\delta m^2_{sol} \ll \delta m^2_{atm} \ll \delta m^2_{LSND}$), 
it is only the largest that actually matters. Therefore for three neutrinos we need only to know
$\delta m^2_{atm} \approx 3\times 10^{-3} eV^2$ \cite{costam1}, whereas for four $\delta m^2_{LSND} \approx
1 eV^2$ \cite{lsnd,costam2}. The second important information is that the spectrum of the masses of light neutrinos
is bounded from above by $m_\nu < 2.5 eV$ \cite{tritium}. With the additional information on the value of one mass,
we recover the complete mass spectrum of neutrinos. 

The basic quantity to consider is the effective neutrino mass, measured in $(\beta\beta)_{0\nu}$ decay
\begin{equation}
\label{basic}
|\langle m_\nu \rangle | \equiv \left| \sum_{i=light} U_{ei}^2 m_i \right| ,
\end{equation}
with $U$ the mixing matrix which we assume
unitary. Current and future bounds on $|\langle m_\nu \rangle |$ are given
in Table~\ref{mnueff}. At present, nothing can be said about the phases of $U$, therefore we
allow them to have the least favorable values. Our analysis is based on the following statement: if the 
minimum of $|\langle m_\nu \rangle |$ with respect to $U_{ei}$ within their allowed 
range, $(|\langle m_\nu \rangle |)_{min}$, exceeds the experimental bound, then neutrinos are Dirac 
particles \cite{future}.
\begin{table}
\caption{present and future bounds on the effective neutrino mass (in $eV$) from 
$(\beta\beta)_{0\nu}$ decay search.}
\begin{tabular}{cr}
\label{mnueff}
$|\langle m_\nu \rangle | <$ & experiment \\
\hline
$.2$ & Heidleberg-Moscow \cite{current} \\
$2\times 10^{-2}$ & GENIUS, first stage, 1yr, 1t \cite{genius} \\
$6\times 10^{-3}$ & GENIUS, second stage, 10yr, 1t \cite{genius} \\
\end{tabular}
\end{table}

Suppose that the solar neutrino oscillations take place mostly between the first and the second 
neutrino states, then \cite{acta}
\begin{equation}
\label{min}
(|\langle m_\nu \rangle |)_{min} = \min \left|l^e m_a - s^e m_b \right|,
\end{equation}
with
\begin{equation}
l^e \in [l^e_-, l^e_+], \;\;\;\; s^e \in [s^e_-, s^e_+],
\end{equation}
and
\begin{equation}
l^e_+ = (1-s^e_+), \;\;\;\; l^e_- = (1-s^e_+) \sqrt{1-\sin^2 2 \theta_{sun}}, 
\end{equation}
where in the case of three neutrinos
\begin{equation}
s^e_+ = s^e_- = |U_{e3}|^2,
\end{equation}
and in the case of four
\begin{equation}
s^e_+ = |U_{e3}|^2+|U_{e4}|^2, \;\;\;\; s^e_- = \left||U_{e3}|^2-|U_{e4}|^2\right|.
\end{equation}
The mass parameters in Eq.~\ref{min} can be chosen in two different ways, depending on the
ordering of masses of the neutrino states.

{\bf Schemes $\boldmath A_3$ and $\boldmath A_4$.} The behavior of the minimum of the
effective mass is the following
\begin{equation}
\label{a3a4eq}
(|\langle m_\nu \rangle |)_{min} = \left\{ \matrix{
s^e_- \sqrt{(m_\nu)_{min}^2+\delta m^2} - l^e_+ (m_\nu)_{min}, \cr
0, \cr
l^e_- (m_\nu)_{min} - s^e_+ \sqrt{(m_\nu)_{min}^2+\delta m^2},
} \right.
\end{equation}
with $\delta m^2 \equiv \delta m^2_{atm}$ for three neutrinos and $\delta m^2 \equiv \delta m^2_{LSND}$
for four. The third range occurs only if $l^e_- > s^e_+$. This can be rewritten as
\begin{equation}
\label{condition}
\sin^2 2 \theta_{sun} < \frac{1-2 s^e_+}{(1-s^e_+)^2},
\end{equation}

Obviously these schemes always allow for a Majorana mass. If however the condition Eq.~\ref{condition} is satisfied, then
it is possible to derive a lower bound on the mass spectrum which excludes this possibility.
This bound will be larger than the value of $(m_\nu)_{min}$ which marks the beginning of the
third range in Eq.~\ref{a3a4eq}
\begin{equation}
(m_\nu)_{min} = \left( \frac{\delta m^2 (s^e_+)^2}{(l^e_-)^2-(s^e_+)^2} \right)^{1/2}.
\end{equation}
This number will be small in the case of three neutrinos if $\sin^2 2\theta_{sun}$ is not too close to one,
due to the small $\delta m^2_{atm}$, but can quickly grow large in the case of four.

{\bf Schemes $\boldmath B_3$ and $\boldmath B_4$.} At the condition Eq.~\ref{condition}
we have
\begin{equation}
(|\langle m_\nu \rangle |)_{min} = l^e_- \sqrt{(m_\nu)_{min}^2+\delta m^2} - s^e_+ (m_\nu)_{min}.
\end{equation}
Obviously, if the masses of the neutrinos are distributed according to these schemes
the neutrino nature can be determined irrespective of the lowest mass, depending solely on 
the experimental bound on $|\langle m_\nu\rangle |$

{\bf Dependence on the mixing matrix.} 
The condition Eq.~\ref{condition} plays a crucial role in the problem. The availability of lower bounds
on the neutrino mass is tightly connected to it. We therefore have to investigate the size of both sides of
the inequality.

\begin{figure}[h]
\epsfig{figure=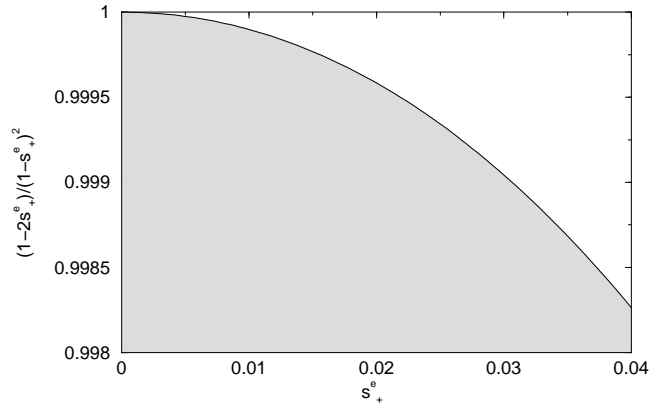, width=8.5cm}
\caption{Allowed region of $\sin^22\theta_{sun}$ by condition Eq.~\ref{condition}.
\label{sinbound}}
\end{figure}

The quantity $s^e_+$ can be fitted to various experimental data. It turns out that the solar neutrino flux
observation is the least constraining since one has a bound of $s^e_+ < .67 - .73$ \cite{solar}. Remark however
that the difference between $\chi^2$ for $s^e_+ = 0$ and the above result is almost irrelevant. Therefore
if other experiments constrain the considered quantity to be close to zero, then solar neutrino oscillation
should be viewed as a two state phenomenon and 3 neutrino fits to solar data alone seem to be
of little value. This will be important subsequently. The currently strongest constraint comes from a combined
analysis of atmospheric and reactor data. Namely one has $s^e_+ < .01 - .02$ \cite{best}. However a recent fit
to the new (830-920 days) data of Super-Kamionkande \cite{nowe} gives a best value of $s^e_+ \simeq .03$.
There has also been some speculation
about strengthening this bound by adding some theoretical assumptions on the shape of the mass matrix. It seems
that $s^e_+ \simeq 0$ is the most preferred value \cite{theor}. In Fig.~\ref{sinbound}, the dependence of the upper
bound on $\sin^2 2\theta_{sun}$ is shown as function of $s^e_+$. In the range $s^e_+ \in [0,.04]$ this bound is
between $.998$ and $1$.

Let us now turn to the experimental bounds on $\sin^2 2\theta_{sun}$. Various fits have been performed for two \cite{two},
three \cite{solar,three} and even four \cite{four} neutrino oscillation scenarios. The most recent which take into account
the full data (rates, energy spectrum, day-night asymmetry in the case of the MSW solution, and the seasonal variation in
the case of the VO solution) give a best fit value of $\sin^2 2\theta_{sun} < 1$. The $95\%c.l.$ bounds for the MSW solution
are given in Table~\ref{zwei}. The most important conclusions are that in neither case do $\sin^2 2\theta_{sun}$
exceed the bound given by condition Eq.~\ref{condition}.

For the Vacuum Oscillation solution best fit value
we take $\sin^2 2\theta_{sun} = .93$ \cite{vo}.
The above values have been used to draw Fig.~\ref{a3a4} and Fig.~\ref{b3b4} for the case $s^e_+ = .01$.

\begin{table}
\caption{Allowed ranges and best fit values of $\sin^22\theta_{sun}$ for the different types of 
oscillations [14]. \label{zwei}}
\begin{tabular}{ccr}
oscillation type & $95\%c.l.$ range & best fit \\
\hline
SMA & $.001 - .01$ & $.0065$ \\
LMA & $.58 - .98$ & $.77$ \\
LOW & $.68 - .98$ & $.90$ \\
\end{tabular}
\end{table}

{\bf Discussion.} The most stringent of the oscillation solutions when question is of the neutrino nature, is
the SMA. Already with the present bound one must adopt the Dirac description in the $B_4$ scheme. With negative results
of GENIUS the same will hold true of the $B_3$ scheme. Currently a bound of $m_\nu > .22eV$ can be taken as a delimiter
of the two neutrino natures for the $A_3, A_4$ and $B_3$ schemes.

At the $95\%c.l.$ the remaining solutions LMA and LOW define the neutrino nature if $m_\nu > 1.5 eV$, with the present bound and
the $A_3, A_4, B_3$ schemes and $m_\nu > 1.1 eV$ in $B_4$. With the GENIUS I bound this can be shifted down to $m_\nu < .16eV$
 ($A_3$), $m_\nu < .14eV$ ($B_3$) and $m_\nu < .22eV$ ($A_4$). Finally GENIUS II
would give $m_\nu < .05eV$ for the $A_3$ and $m_\nu < .12eV$ for the $A_4$ scheme, but would completely exclude a Majorana neutrino in the
$B_3$ scheme.

We do not have a $95\%c.l.$ bound for the VO solution. However, with the assumption that here also $\sin^22\theta_{sun} < .98$ the same
conclusions would hold as above.

If one takes the best fit values for the different solutions, all of the bounds can be improved.

An interesting consideration is to take into account the current best fit for the Hot Dark Matter neutrino constituent.
One then has $\sum_{light} m_\nu \simeq 2eV$ \cite{hdm}. The SMA solution rules out in this case Majorana neutrinos in the $A_3$ and
$B_3$ schemes. With a negative result from GENIUS all solutions share this property in the $A_3, B_3$ and $B_4$ schemes.
Obviously, if we take the previous result $\sum_{light} m_\nu \simeq 4-5eV$ \cite{previous}, then the constraint is even stronger
as already now most of the schemes are ruled out and GENIUS would close the situtation.

{\bf Final considerations.} Several points must be stressed. First, the bound on $|\langle m_\nu \rangle |$ depends strongly
on the precise determination of nuclear matrix elements. It is acknowledged that they suffer from an imprecision at the
level of a factor of 3 \cite{pdg}. Second, the confidence level that we assumed, $95\%$ corresponds to $2\sigma$. At $3\sigma$ which 
is $99\%c.l.$, the $\sin^22\theta_{sun}$ value of one is allowed. This unfortunately ruins the validity of condition
Eq.~\ref{condition} with the same effect on our predictions. Third, the figures have been drawn for $\delta m^2_{LSND} = 1eV^2$.
The experimentally allowed range is $\delta m^2_{LSND} = .2 - 2 eV^2$. Adoption of some other difference of mass squared would
require a rescaling of the results for the $A_4$ and $B_4$ neutrino schemes.

{\bf Conclusions.} The nature of neutrinos that seemed so evasive is closer and closer to the reach of experiments.
Combining oscillation data, tritium beta decay and neutrinoless double beta decay is a powerful method of constraining
the parameter space in which Majorana neutrinos are allowed. In this paper we have derived bounds on the masses in several
schemes and shown which schemes are or will be excluded out of consideration if neutrinos are to be truly neutral particles.
We still have to wait for MiniBoone or SNO, to decide on the existence of the sterile neutrino, and for SNO, Borexino, Kamland
to determine the solar neutrino solution and finally for more statistics in all the experiments to constrain more precisely the $\sin^22\theta_{sun}$.

{\bf Acknowledgments.} This work was supported by the Polish Committee for Scientific Research under Grant No. 2P03B05418.
J.G. would like to thank the Alexander von Humboldt-Stiftung for fellowship.

\end{document}